\documentclass{article}
\usepackage{pgf}
\usepackage{xcolor}

\usepackage{PRIMEarxiv}
\usepackage[utf8]{inputenc} 
\usepackage[T1]{fontenc}    
\usepackage{hyperref}       
\usepackage{url}            
\usepackage{booktabs}       
\usepackage{amsfonts}       
\usepackage{nicefrac}       
\usepackage{microtype}      
\usepackage{lipsum}         
\usepackage{fancyhdr}       
\usepackage{graphicx}       
\usepackage{caption}
\usepackage{enumitem}
\usepackage{array}
\usepackage{float}
\usepackage{ragged2e}
\usepackage{multirow} 
\usepackage{colortbl}  
\usepackage{float}

\graphicspath{{media/}}     
\usepackage[utf8]{inputenc}

\usepackage{colortbl}       
\usepackage{pgfplots}       
\usepackage{pgfplotstable}  

\newfloat{exhibit}{htbp}{loe}
\floatname{exhibit}{Exhibit}

\title{Measuring Determinism in Large Language Models\\
       for Software Code Review}

\author{
  Eugene Klishevich \\
   \And
  Yegor Denisov-Blanch$^1$ \\
   \And
  Simon Obstbaum \\
   \And
  Igor Ciobanu \\
   \And
  Michal Kosinski$^1$ \\
}

\begin{document}
\maketitle

\vspace{-4em}
\begin{center}
  $^1$Stanford University
\end{center}
\vspace{1.5em}

\begin{abstract}
Large Language Models (LLMs) promise to streamline software code reviews, but their ability to produce consistent assessments remains an open question. In this study, we tested four leading LLMs---GPT-4o mini, GPT-4o, Claude 3.5 Sonnet, and LLaMA 3.2 90B Vision---on 70 Java commits from both private and public repositories. By setting each model’s temperature to zero, clearing context, and repeating the exact same prompts five times, we measured how consistently each model generated code-review assessments. Our results reveal that even with temperature minimized, LLM responses varied to different degrees. These findings highlight a consideration about the inherently limited consistency (test-retest reliability) of LLMs---even when the temperature is set to zero---and the need for caution when using LLM-generated code reviews to make real-world decisions.

\end{abstract}

\section{Introduction}
Large Language Models (LLMs) have emerged as transformative tools across many domains, including software development. One area where they can be particularly helpful is code review---the process of examining changes in code to maintain development quality. Although code review is vital to ensure code reliability and facilitate future development~\cite{jetbrains2024}, it often consumes significant time and can be tedious for developers~\cite{silverthorne2021,yang2024}, which explains growing interest in automating or assisting this task with LLMs~\cite{stackoverflow2024,cihan2024}.

For LLMs to be suitable for code review, they must produce consistent and accurate feedback when faced with similar or identical code changes. This consistency is critical: a model that evaluates the same (or nearly the same) code differently can call into question the viability of automated code review assistance. Yet, even when the “temperature” (the parameter that controls randomness) is set to zero, LLMs often deliver varied assessments. Our study investigates the degree of determinism exhibited by different LLMs in these zero-temperature scenarios, focusing on GPT-4o, GPT-4o mini, Claude, and LLaMA.

Our findings highlight that, despite minimizing randomness, LLMs can still produce different outputs for identical inputs. This result points to important factors beyond temperature that may influence a model’s behavior. Researchers and practitioners must address these sources of variation to ensure LLM-powered code reviews yield reliable, reproducible results.

\section{Methodology}

\subsection{Experimental Setup Overview}
We used four LLMs (Section~\ref{sec:llmselection}), each tasked with assessing seventy code commits (Section~\ref{sec:dataacquisition}) by answering five questions (Section~\ref{sec:questions}). We tested three different prompt lengths (Section~\ref{sec:promptdesign}) and, for each prompt length, ran five independent runs with temperature set to zero and a cleared context window. Finally, we evaluated the outputs for consistency by comparing results from identical inputs (Section~\ref{sec:measuringconsistency}).

In total, we gathered 21,000 responses (4 LLMs $\times$ 3 prompt lengths $\times$ 5 questions $\times$ 70 commits $\times$ 5 runs), supporting the statistical significance of our findings (Section~\ref{sec:stats}).

\subsection{LLM Selection}\label{sec:llmselection}
We used four state-of-the-art models---GPT-4o mini, GPT-4o, Claude 3.5 Sonnet, and Llama 3.2 90B Vision, which top benchmarks such as Aider’s Leaderboards (Aider, 2024) and SweBench (SweBench Consortium, 2024).

\subsection{Data Acquisition and Commit Selection}\label{sec:dataacquisition}
We used 70 code commits chosen from a larger dataset of 1.73 million commits gathered in a previous study~\cite{denisovblanch2024}. These commits came from both private repositories (collected from 108 organizations recruited via LinkedIn), at a ratio of one public commit for every five private commits. This ensured diversity in commit size and content. We focused on Java code because of its widespread use across diverse application domains~\cite{stackoverflow2024}.

\subsection{Prompt Design}\label{sec:promptdesign}
Drawing on Chen et al.~\cite{chen2024} and Sahoo et al.~\cite{sahoo2024}, we structured each query to the LLMs in three parts:

\begin{enumerate}
    \item \textbf{Instructional Brief:} A short text describing the LLM’s role and main instructions (e.g., “You are a senior software engineer reviewing a code commit…”).
    \item \textbf{Code Snippet:} The entire patch file from one of the 70 selected commits, including every added, modified, and deleted line of code.
    \item \textbf{Question Prompt:} One of five code-review questions (Q1--Q5).
\end{enumerate}

We tested three variations of prompt length to see whether the level of detail affects LLM responses:
\begin{itemize}
    \item \textbf{Simple (A)} (\(\sim\)100 words)
    \item \textbf{Balanced (B)} (\(\sim\)350 words)
    \item \textbf{Explanatory (C)} (\(\sim\)750 words)
\end{itemize}

\begin{table}[htbp]
\centering
\small
\caption{Approximate Word Counts for Each Part of the Prompt}
\label{tab:promptlength}
\begin{tabular}{lccccccc}
\toprule
\textbf{Prompt Type} & \textbf{Avg. Total} & \textbf{Instructional Brief} & \textbf{Q1} & \textbf{Q2} & \textbf{Q3} & \textbf{Q4} & \textbf{Q5}\\
\midrule
Simple (A)      & 99  & 36  & 71  & 78  & 64  & 53  & 50  \\
Balanced (B)    & 340 & 158 & 184 & 153 & 179 & 184 & 212 \\
Explanatory (C) & 750 & 390 & 256 & 363 & 388 & 330 & 479 \\
\bottomrule
\end{tabular}
\end{table}

\subsection{Questions for Commit Code Assessment}\label{sec:questions}
To evaluate each commit we asked questions covering five topics (Q1--Q5), summarized in 
Table~\ref{tab:commit-evaluation}. Note that the text in the “Question Topic” column is a 
concise representation; in practice, each question was framed in a more extensive prompt 
(see Appendix~\ref{sec:appendixprompts} for the complete prompt text). Questions~1 and~2 used a Fibonacci 
scale for their response options, a common Agile practice that helps distinguish task sizes 
more effectively than a simple linear scale.

\begin{table}[htbp]
\centering
\small
\caption{Commit Evaluation Questionnaire}
\label{tab:commit-evaluation}
\begin{tabular}{
  >{\raggedright\arraybackslash}p{0.05\textwidth}
  >{\RaggedRight\arraybackslash}p{0.61\textwidth}
  >{\RaggedRight\arraybackslash}p{0.28\textwidth}
}
\toprule
\textbf{No.} & \textbf{Question Topic} & \textbf{Response Options} \\
\midrule
Q1 
  & How many hours would it take you to just write the code in this commit? 
  & 1, 2, 3, 5, 8, 13, 21, 34, 55, 89 \\
\midrule
Q2 
  & How many hours would it take to fully implement this commit (including debugging \& QA)? 
  & 1, 2, 3, 5, 8, 13, 21, 34, 55, 89 \\
\midrule
Q3 
  & What is the experience level of the author?
  & \begin{tabular}[t]{@{}l@{}}%
       Novice/Beginner\\
       Basic/Elementary\\
       Intermediate\\
       Advanced\\
       Expert/Master
    \end{tabular} \\
\midrule
Q4 
  & How difficult is the problem that this commit solves? 
  & \begin{tabular}[t]{@{}l@{}}%
       Very Easy\\
       Easy\\
       Moderate\\
       Challenging\\
       Very Challenging
    \end{tabular} \\
\midrule
Q5 
  & How maintainable is this commit? 
  & \begin{tabular}[t]{@{}l@{}}%
       Poor\\
       Below Average\\
       Average\\
       Good\\
       Excellent
    \end{tabular} \\
\bottomrule
\end{tabular}
\end{table}

\subsection{Measuring Consistency}\label{sec:measuringconsistency}
To obtain discrete responses from the set of options in Table~\ref{tab:commit-evaluation}, we instructed the LLMs to include a standardized JSON object in their replies (see Appendix: Prompts). This structured format allowed us to systematically record each response. We then used Pearson’s correlation coefficient to measure the consistency of the recorded outputs. For each combination of model and prompt type, we calculated the correlation of answers across the five runs for each question. We then averaged these correlations across all five questions to obtain an overall consistency score per model-prompt pair. Finally, we pooled these scores across the three prompt types to summarize each model’s overall consistency. To estimate the reliability of our correlation values, we computed 95\% confidence intervals via bootstrap sampling (10,000 resamples).

\subsection{Statistical Significance and Sample Size}\label{sec:stats}
We used a permutation test (\(p < 0.05\)) to confirm that our sample size of 70 commits, each evaluated five times for five questions, was sufficient to detect statistically significant differences in model consistency.

\section{Results}

\subsection{Model Determinism Measurement}
The consistency of outputs from four LLMs is demonstrated in Table~\ref{tab:determinism}.

\begin{table}[H]
    \centering
    \caption{Model Determinism Measured by Correlation Across Repeated Runs}
    \label{tab:determinism}
    \centering
    \includegraphics[width=\textwidth]{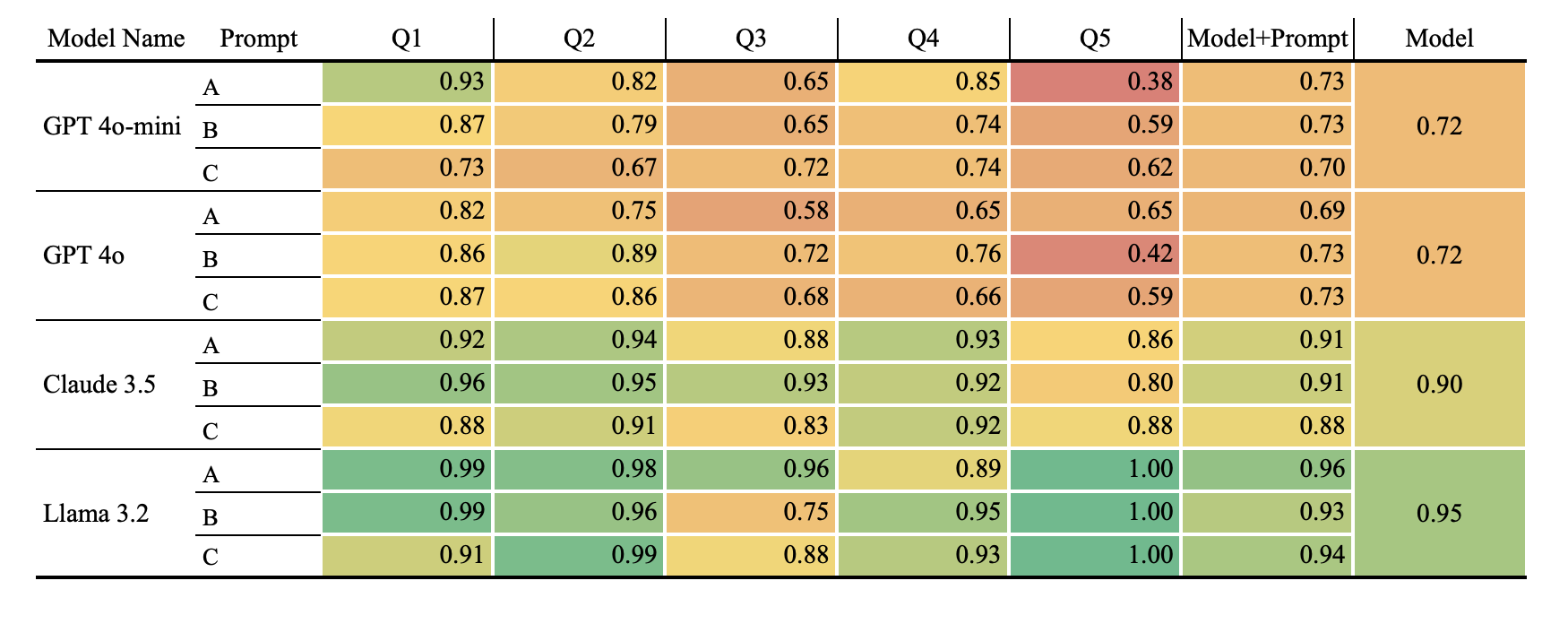}
\end{table}

Exhibit~\ref{ex:observed-correlation} compares model correlations (the data from the column “Model” of Table~\ref{tab:determinism}), and a 95\% confidence interval for the correlation of each model.

\begin{exhibit}[htbp]
    \caption{Observed Correlation and Confidence Intervals}
    \label{ex:observed-correlation}
    \centering
    \includegraphics[width=\textwidth]{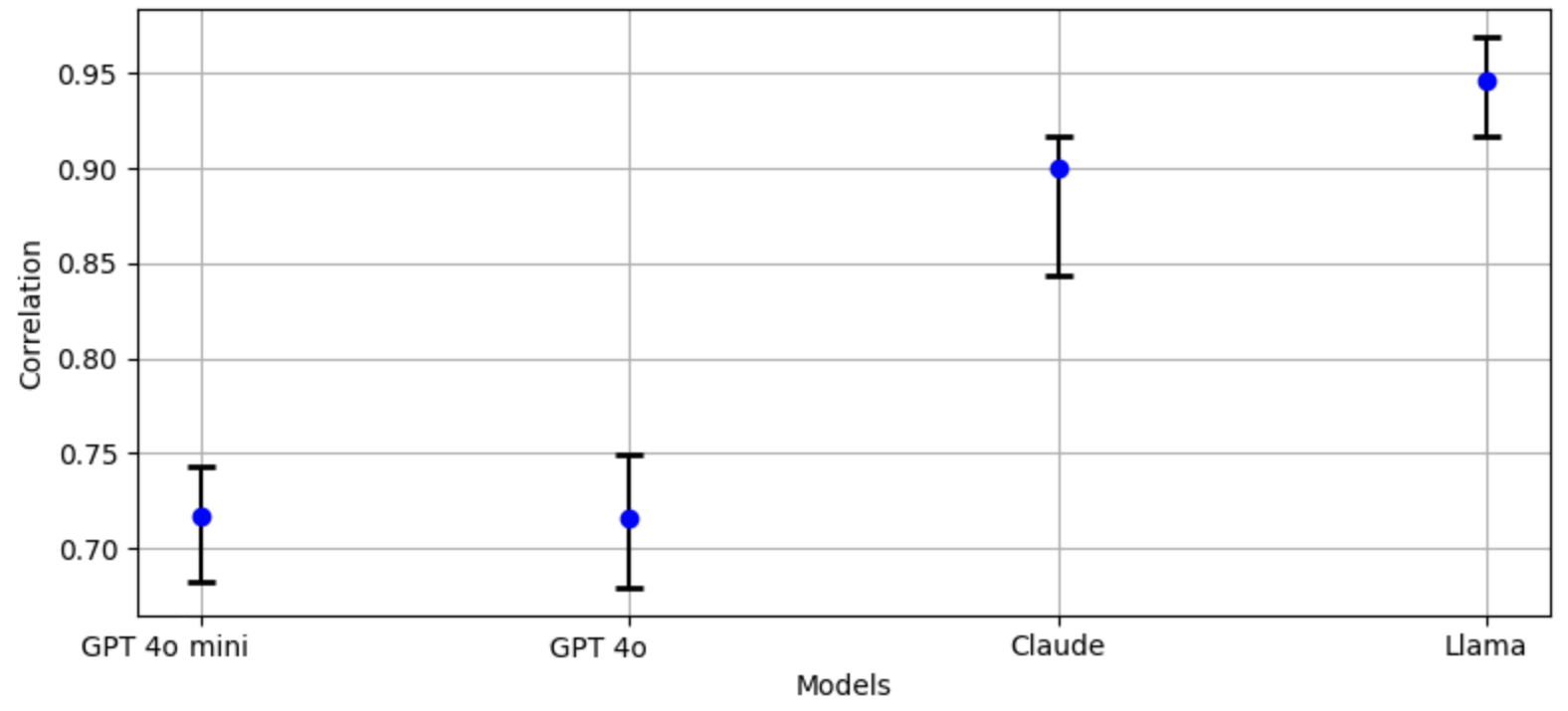}
\end{exhibit}

\subsection{Comparison with Human Reviewer Consistency}

Although LLM inconsistencies may raise concerns, it is useful to compare them (Table~\ref{tab:humancompare}) against the current standard for code assessments: human reviewers. Prior work~\cite{denisovblanch2024} shows that multiple expert raters can also display notable variability, as measured by ICC\(_{2,k}\). This implies that LLMs’ reliability may be comparable to that of human reviewers, though additional studies are needed to confirm whether LLM variability is higher, lower, or simply different in nature.

\begin{table}[H]
\centering
\caption{Human Raters' ICC\(_{2,k}\) vs. LLM Determinism}
\label{tab:humancompare}
\begin{tabular}{clccccc}
\toprule
\textbf{No.} & \textbf{Question Topic} & \textbf{Human ICC\(_{2,k}\)} & 
\textbf{GPT-4o mini} & \textbf{GPT-4o} & \textbf{Claude 3.5} & \textbf{Llama 3.2}\\
\midrule
Q1 & Hours to just write code & 0.81 & 0.84 & 0.85 & 0.92 & 0.96 \\
Q2 & Hours to implement       & 0.82 & 0.76 & 0.83 & 0.93 & 0.98 \\
Q3 & Experience level         & 0.61 & 0.67 & 0.66 & 0.88 & 0.86 \\
Q4 & Difficulty               & 0.78 & 0.78 & 0.69 & 0.92 & 0.92 \\
Q5 & Maintainability          & 0.52 & 0.53 & 0.55 & 0.85 & 1.00 \\
\bottomrule
\end{tabular}
\end{table}

\section{Discussion}
Our findings confirm that Large Language Models (LLMs) exhibit inherent non‐determinism, even with the temperature set to zero. While earlier studies have noted that LLMs can produce varied outputs under identical prompts, our work extends this understanding by quantifying the extent of these variations and directly comparing the determinism of four prominent LLMs in the context of code review. Notably, the degree of determinism follows a clear gradient: GPT‐4o mini and GPT‐4o show lower consistency, Claude provides moderate consistency, and Llama demonstrates the highest determinism among the tested models.

\subsection{Understanding LLM Inconsistency}
Our research demonstrates that LLMs can exhibit varying consistency in code review tasks, even when the temperature parameter is set to zero---underlining that “zero temperature” does not guarantee fully deterministic outputs. While temperature indeed affects determinism, other model‐specific factors, such as floating‐point precision, token sampling quirks, architecture, and training data~\cite{atil2024}), also play significant roles. Moreover, some closed‐source LLMs may deliberately limit determinism to safeguard proprietary techniques from reverse engineering.

\subsection{Practical Implications}
Excessive variability in AI‐assisted code reviews can undermine the assessment process and impede wider adoption of these tools. One stopgap approach is to run multiple LLM iterations and either average their outputs or select the most frequent answer---though this significantly increases computational cost. Other methods include fine‐tuning, inference‐time techniques, model ensembles, and other training strategies to reduce variability~\cite{pornprasit2024,yu2024,chang2024,singh2024}. Additionally, open‐source models allow deeper visibility into the factors that drive inconsistency, offering a promising avenue for developing more predictable, stable code review assistance.

\subsection{Limitations and Future Research}
While our findings are promising, several constraints should be noted. First, the dataset comprised 70 commits, which, although representative of real-world patterns, limits the breadth of our conclusions. Second, all commits were sourced from Java-based projects, potentially restricting generalizability to other programming paradigms. Third, because maintainability is subjective and depends on human judgment, low inter-rater reliability (ICC\(_{2,k}\)) inherently limits how closely any model can align with human evaluations. Future work should explore larger datasets, additional languages, refined methods to handle subjectivity in code reviews, and newer models (e.g. DeepSeek R1, and GPT o1 and other advanced reasoning models) as well as fine-tuned variants (e.g. Cursor, Github Copilot) and explore advanced techniques for reducing variability. Together with the measure of consistency (as a measure of precision) of LLMs performing software code review, another important parameter to consider is validity (the level of accuracy in responses), which we plan to explore in our next paper.

\section{Conclusion}
Our study measured the actual level of determinism of LLMs when applied to software code review tasks. The evaluation not only demonstrates the level of consistency for each model, but also provides an opportunity to compare this consistency across assessed LLMs. While LLMs show promise for automating code reviews, their non-deterministic nature and limited consistency, even under low-temperature settings, present challenges. Both researchers and practitioners must consider these limitations and implement strategies to mitigate risk of errors when integrating LLMs into software development workflows.

\vspace{1em}

\bibliographystyle{unsrt}
\bibliography{references}

\bigskip
\appendix
\section{Appendix: Prompts}
\label{sec:appendixprompts}

\textbf{Appendix: Prompts}

\subsubsection{Instructional Brief A – Simple}

You are a highly experienced Java developer, tech lead, and software architect. Your task is to evaluate a Java code commit made to a codebase. Carefully and thoroughly analyze the code changes introduced in this commit.

\paragraph{Q1 Prompt}

Estimate the number of hours it would take to write this code from scratch, assuming you could fully focus on this task without interruptions.

Select the most appropriate estimate (in hours) from the following options: 1, 2, 3, 5, 8, 13, 21, 34, 55, 89.

Then, generate a JSON object with the following structure:

\{

"commit\_hash": "insert commit hash here", 

"MP": "1",

"Q": "1",

"Answer": "insert estimated value that was selected earlier"

\}

\paragraph{Q2 Prompt}

Estimate the number of hours it would take to implement this commit from start to finish, including debugging and QA cycles, assuming you could fully focus on this task without interruptions.

Select the most appropriate estimate (in hours) from the following options: 1, 2, 3, 5, 8, 13, 21, 34, 55, 89.

Then, generate a JSON object with the following structure:

\{

"commit\_hash": "insert commit hash here", 

"MP": "1",

"Q": "2",

"Answer": "insert estimated value that was selected earlier"

\}

\paragraph{Q3 Prompt}

Based on the code in this commit, determine the experience level of the code author. 

Select the most appropriate level from the following experience hierarchy, ranked from lowest to highest: Novice/Beginner, Basic/Elementary, Intermediate, Advanced, Expert/Master.

Then, generate a JSON object with the following structure:

\{

"commit\_hash": "insert commit hash here", 

"MP": "1",

"Q": "3",

"Answer": "insert estimated value that was selected earlier"

\}

\paragraph{Q4 Prompt}

Assess the difficulty of the problem that this commit addresses. 

Select the most appropriate difficulty level from the following options: Very Easy, Easy, Moderate, Challenging, Very Challenging.

Then, generate a JSON object with the following structure:

\{

"commit\_hash": "insert commit hash here", 

"MP": "1",

"Q": "4",

"Answer": "insert estimated value that was selected earlier"

\}

\paragraph{Q5 Prompt}

Evaluate the maintainability of the code in this commit.

Select the most appropriate rating from the following options: Poor, Below Average, Average, Good, Excellent.

Then, generate a JSON object with the following structure:

\{

"commit\_hash": "insert commit hash here", 

"MP": "1",

"Q": "5",

"Answer": "insert estimated value that was selected earlier"

\}

\subsubsection{Instructional Brief B – Balanced}

You are a highly experienced Java software engineer with a strong background in software development, code review, and architectural design. Your task is to analyze a Java code commit, presented in a diff format, which includes additions, modifications, and deletions compared to the previous version.

In your evaluation, you should:

Examine the purpose and functionality of the changes based on the provided code.

Assess the code's structure, organization, and adherence to object-oriented principles such as encapsulation, inheritance, and abstraction.

Evaluate the complexity, readability, and clarity of the code, including naming conventions and commenting.

Consider the use of data structures, algorithms, and their potential impact on performance and maintainability.

Review any dependencies or integrations introduced in the code, assessing their compatibility and impact on maintainability.

Your analysis should focus on providing clear and objective insights based solely on the code provided in the commit. Avoid assumptions about parts of the codebase not included in the commit or external documentation.

\paragraph{Q1 Prompt}

Estimate the number of hours it would take to write the code changes in this commit from scratch, assuming you could fully focus on this task without interruptions. Focus solely on the time required to write the code, excluding activities such as debugging, testing, or quality assurance.

To arrive at your estimate, consider:

Complexity: Evaluate the algorithms, data structures, and design patterns used and their impact on implementation time.

Volume: Assess the amount of code added, modified, or deleted and the effort required to write it.

Domain Familiarity: Assume familiarity with the technologies and problem domain relevant to the code.

Organization: Account for the time required to organize and structure the code, such as creating classes, methods, or interfaces.

Best Practices: Reflect on the time needed to ensure adherence to coding standards and best practices.

Briefly explain your reasoning and select the most appropriate estimate (in hours) from: 1, 2, 3, 5, 8, 13, 21, 34, 55, 89.

Then, generate a JSON object with the following structure:

\{

"commit\_hash": "insert commit hash here", 

"MP": "B",

"Q": "1",

"Answer": "insert estimated value that was selected earlier"

\}

\paragraph{Q2 Prompt}

Estimate the number of hours it would take to implement the code changes in this commit from start to finish, including debugging, testing, and quality assurance, assuming uninterrupted focus. Your estimate should cover all work needed to ensure the code is fully functional, reliable, and deployable.

Consider the following:

Complexity: How the algorithms, data structures, and patterns used affect effort.

Volume: The size of the code changes and time to verify correctness.

Debugging and Testing: Time for fixing bugs, running tests, and ensuring quality.

Familiarity: Assume familiarity with the domain and technologies involved.

Best Practices: Time to follow standards, write documentation, and ensure maintainability.

Briefly explain your reasoning and select the most appropriate estimate (in hours) from: 1, 2, 3, 5, 8, 13, 21, 34, 55, 89.

Then, generate a JSON object with the following structure:

\{

"commit\_hash": "insert commit hash here", 

"MP": "B",

"Q": "2",

"Answer": "insert estimated value that was selected earlier"

\}

\paragraph{Q3 Prompt}

Based on the code in this commit, assess the experience level of the code author in Java programming and software development. Use the provided code to evaluate their expertise. When making your assessment, consider:

Code Quality and Style: Adherence to coding standards, formatting, and naming conventions.

Use of Language Features: Effective use of Java features such as generics, streams, or annotations.

Object-Oriented Principles: Application of encapsulation, inheritance, and abstraction.

Design Patterns: Use and appropriateness of design patterns like Factory or Singleton.

Structure and Organization: Logical organization of classes, interfaces, and methods.

Error Handling: Robustness in managing exceptions, edge cases, and unexpected inputs.

Best Practices: Alignment with principles like SOLID, DRY, and KISS, and the presence of clear comments or documentation.

Complexity and Efficiency: Optimization and scalability of algorithms and data structures.

Briefly explain your reasoning and select the most appropriate level from the following hierarchy: Novice/Beginner, Basic/Elementary, Intermediate, Advanced, Expert/Master.

Then, generate a JSON object with the following structure:

\{

"commit\_hash": "insert commit hash here", 

"MP": "B",

"Q": "3",

"Answer": "insert estimated value that was selected earlier"

\}

\paragraph{Q4 Prompt}

Assess the difficulty level of the problem addressed by the code changes in this commit. Focus on the complexity and challenges associated with the problem itself, not the effort required to implement the solution. In your assessment, consider:

Nature of the Problem: The goal of the changes and whether they involve straightforward or complex functionality.

Complexity of Requirements: The intricacy of the implied requirements, such as complex logic, algorithms, or data processing.

Technical Challenges: Any hurdles like concurrency, optimization, or resource management.

Dependencies and Interactions: Integration with existing systems, modules, or external libraries.

Scope and Impact: The extent of changes required and their potential impact on the overall system.

Advanced Concepts: Use of advanced programming techniques or design patterns that indicate higher difficulty.

Potential Risks: Risks or challenges related to security, data integrity, or unforeseen obstacles.

Briefly explain your reasoning and select the most appropriate difficulty level from the following options: Very Easy, Easy, Moderate, Challenging, Very Challenging.

Then, generate a JSON object with the following structure:

\{

"commit\_hash": "insert commit hash here", 

"MP": "B",

"Q": "4",

"Answer": "insert estimated value that was selected earlier"

\}

\paragraph{Q5 Prompt}

Evaluate the maintainability of the code changes in this commit. Based solely on the provided code, assess how easy it would be for developers to understand, modify, fix, and extend the code in the future. In your evaluation, consider:

Readability: Clarity of naming conventions, formatting, and use of comments or self-explanatory code.

Organization and Structure: Logical arrangement of classes, methods, and packages, and the modularity of components.

Coding Standards and Best Practices: Consistency with coding conventions and principles like SOLID, DRY, and KISS.

Complexity: Simplicity versus unnecessary complexity, and the code's cyclomatic or algorithmic complexity.

Reusability: Use of design patterns and the potential for easy extension or modification.

Error Handling: Robustness in handling exceptions and edge cases, with proper logging or reporting mechanisms.

Dependencies: Management of external libraries, APIs, or modules, and their impact on maintainability.

Testability: Ease of writing and executing unit or integration tests, and whether the code structure facilitates testing.

Documentation: Presence and quality of comments or documentation explaining complex sections of code.

Briefly explain your reasoning and select the most appropriate rating from the following options: Poor, Below Average, Average, Good, Excellent.

Then, generate a JSON object with the following structure:

\{

"commit\_hash": "insert commit hash here", 

"MP": "B",

"Q": "5",

"Answer": "insert estimated value that was selected earlier"

\}

\subsubsection{Instructional Brief C – Explanatory}

You are a highly experienced Java software engineer with over 10 years of professional experience in software development, code review, and architectural design. Your background includes extensive work with object-oriented programming, design patterns, software development methodologies, and a deep understanding of industry best practices. 

You have been provided with a code commit that includes Java code changes presented in a diff format (showing additions and deletions compared to the previous version). Start by performing a comprehensive and meticulous analysis of these code changes.

In your analysis, you should:

\textbf{Thoroughly Examine the Code Changes:}

Review all additions, modifications, and deletions in the code.

 Understand the purpose and functionality of the changes.

 Consider how these changes fit within the likely context of the application, based solely on the provided code.

\textbf{Assess Code Structure and Organization:}

Evaluate the arrangement and relationships of classes, interfaces, methods, and other structural elements.

Analyze the use of object-oriented principles such as encapsulation, inheritance, polymorphism, and abstraction.

\textbf{Evaluate Complexity and Readability:}

Analyze the cyclomatic and n-path complexity, control flow, and data flow of the code.

Assess the readability and clarity of the code, including naming conventions and commenting.

Determine whether the code introduces unnecessary complexity or technical debt.

\textbf{Analyze Use of Data Structures and Algorithms:}

Examine the selection and implementation of data structures (e.g., lists, maps, sets, queues).

Assess the efficiency and suitability of algorithms used.

Consider the impact on performance, scalability, and resource utilization.

\textbf{Consider Dependencies and External Integrations:}

Identify any external libraries, frameworks, or APIs used in the code.

Evaluate how these dependencies affect the code's portability, maintainability, and compatibility.

\textbf{Review Error Handling and Exception Management:}

Assess how the code handles exceptions, errors, and edge cases.

Evaluate the robustness and reliability of the code in handling unexpected inputs or conditions.

\textbf{Evaluate Testing and Quality Assurance Considerations:}

Consider how testable the code is based on its structure and dependencies.

\textbf{Assess Maintainability and Extensibility:}

Determine how easily the code can be understood, maintained, and modified by other developers.

Evaluate the modularity and reusability of the code components.

Your analysis should be objective, detailed, and based solely on the code provided in the commit. Do not make assumptions about external documentation, specifications, or parts of the codebase not included in the commit. Focus on providing insights and observations that reflect your depth of experience in Java software engineering. 

\paragraph{Q1 Prompt}

Your task is to estimate the number of \textbf{hours} it would take you to write the code changes in this commit from scratch, assuming you could fully focus on this task without any interruptions. This estimation should include only the time required to write the code itself, excluding any time for debugging, testing, or quality assurance activities.

To provide a thorough and precise estimation, please also carefully consider the following aspects:

\textbf{Complexity of Implementation:} Assess the complexity of the algorithms, data structures, and design patterns used (if applicable). Determine how these factors would affect the time needed to implement the code.

\textbf{Volume of Code Changes:} Thoroughly evaluate the content of code added, modified, or deleted. Consider the effort required to produce this code.

\textbf{Familiarity with the Domain:} Assume you are familiar with the problem domain and the technologies involved.

\textbf{Code Structure and Organization:} Consider the time needed to design and organize code structures such as classes, interfaces, and methods effectively.

\textbf{Dependencies and Integrations:} Evaluate cross-boundary and circular dependencies

\textbf{Use of Best Practices:} Reflect on the time required to implement the code following industry best practices and coding standards.

After analyzing these factors, provide an explanation for your estimation, highlighting the factors that influenced your decision.

Next, select the most appropriate time estimate (in hours) from the following set of values: \textbf{1, 2, 3, 5, 8, 13, 21, 34, 55, 89}

Then, generate a JSON object with the following structure:

\{

"commit\_hash": "insert commit hash here", 

"MP": "C",

"Q": "1",

"Answer": "insert estimated value that was selected earlier"

\}

\paragraph{Q2 Prompt}

Your task is to estimate the number of \textbf{hours} it would take you to implement the code changes in this commit \textbf{from start to finish}, including debugging, testing, and quality assurance activities, assuming you could fully focus on this task without any interruptions. This estimation should encompass all the work required to ensure the code is fully functional, reliable, and ready for deployment.

To provide a thorough and precise estimation, please also carefully consider the following aspects:

\textbf{Complexity of Implementation:} Assess the complexity of the algorithms, data structures, and design patterns used. Determine how these factors would affect the time needed to implement, debug, and test the code thoroughly.

\textbf{Volume of Code Changes:} Thoroughly evaluate the content of code added, modified, or deleted. Consider the effort required not only to write this amount of code but also to verify its correctness and performance.

\textbf{Debugging and Testing Effort:} Estimate the time required to identify and fix potential bugs, write unit tests, perform integration testing, and conduct any necessary quality assurance cycles. Consider the likelihood of issues arising based on the code's complexity.

\textbf{Familiarity with the Domain:} Assume you are familiar with the problem domain and the technologies involved. Consider how this familiarity would influence both the implementation and the debugging/testing phases.

\textbf{Code Structure and Organization:} Consider the time needed to design and organize code structures such as classes, interfaces, and methods effectively, as well as to ensure they are testable and maintainable.

\textbf{Dependencies and Integrations:} Factor in the effort required to integrate any external libraries, APIs, or dependencies present in the code including cross boundary and circular.

\textbf{Use of Best Practices:} Reflect on the time required to implement the code following industry best practices and coding standards, including thorough documentation and adherence to coding conventions.

After analyzing these factors, provide an explanation for your estimation, highlighting the factors that influenced your decision.

Next, select the most appropriate time estimate (in hours) from the following set of values: \textbf{1, 2, 3, 5, 8, 13, 21, 34, 55, 89}

Then, generate a JSON object with the following structure:

\{

"commit\_hash": "insert commit hash here", 

"MP": "C",

"Q": "2",

"Answer": "insert estimated value that was selected earlier"

\}

\paragraph{Q3 Prompt}

Your task is to estimate the \textbf{experience level} of the author of the code changes in this commit. Based solely on the provided code, assess the author's experience in Java programming and software development.

To make a thorough and accurate assessment, please carefully consider the following aspects:

\textbf{Code Quality and Style:} Evaluate the quality of the code, including adherence to coding standards, consistency in formatting, and use of appropriate naming conventions for classes, methods, and variables.

\textbf{Use of Language Features:} Assess the author's utilization of Java language features, such as generics, lambda expressions, streams, annotations, and other advanced constructs. Consider how effectively these features are employed.

\textbf{Application of Object-Oriented Principles:} Examine the implementation of object-oriented principles, including encapsulation, inheritance, polymorphism, and abstraction. Determine how well the author applies these concepts to create a robust and maintainable codebase.

\textbf{Implementation of Design Patterns:} Identify any design patterns used (e.g., Builder, Factory, Singleton, Strategy, Observer) and evaluate whether they are appropriately and effectively applied.

\textbf{Code Structure and Organization:} Consider the modularity, cohesiveness, and organization of the code. Evaluate the effective use of classes, interfaces, methods, and packages to structure the application logically.

\textbf{Error Handling and Exception Management:} Assess how the author handles exceptions, errors, and edge cases. Evaluate the robustness and reliability of the code in managing unexpected inputs or conditions.

\textbf{Use of Best Practices:} Reflect on the adherence to industry best practices and coding standards, such as the SOLID principles, DRY (Don't Repeat Yourself), and KISS (Keep It Simple, Stupid). Consider the presence of code comments and documentation where appropriate.

\textbf{Complexity and Efficiency:} Analyze the complexity of the code and the efficiency of algorithms and data structures used. Determine whether the author has optimized the code for performance and scalability.

\textbf{Innovativeness and Problem-Solving Skills:} Consider any innovative solutions or clever problem-solving approaches demonstrated in the code. Assess how the author tackles challenges and implements effective solutions.

After analyzing these factors, provide an explanation for your estimation, highlighting the key indicators that influenced your determination of the author's experience level.

Next, select the most appropriate level from the following experience hierarchy, ranked from lowest to highest: \textbf{Novice/Beginner, Basic/Elementary, Intermediate, Advanced, Expert/Master}

Then, generate a JSON object with the following structure:

\{

"commit\_hash": "insert commit hash here", 

"MP": "C",

"Q": "3",

"Answer": "insert estimated value that was selected earlier"

\}

\paragraph{Q4 Prompt}

Your task is to estimate the \textbf{difficulty level} of the problem that the code changes in this commit aim to solve. Based solely on the provided code, evaluate the complexity and challenges associated with the problem itself, not the implementation effort.

To make a thorough and accurate assessment, please carefully consider the following aspects:

\textbf{Nature of the Problem:}

Determine the objective of the code changes and the specific problem they address.

Consider whether the problem involves straightforward functionality or complex features.

\textbf{Complexity of Requirements:}

Assess the intricacy of the requirements implied by the code.

Evaluate if the problem requires handling complex logic, algorithms, or data processing.

\textbf{Technical Challenges:}

Identify any technical hurdles that need to be overcome, such as concurrency, synchronization, performance optimization, or resource management.

Consider the necessity of advanced programming techniques or specialized knowledge.

\textbf{Dependencies and Interactions:}

Examine how the problem interacts with existing systems, modules, or external dependencies.

Assess the complexity introduced by integrating with APIs, libraries, or third-party services.

\textbf{Scope and Impact:}

Evaluate the extent of the changes required to solve the problem, including the number of components or layers affected.

Consider the potential impact on the overall system, such as critical functionality or high-stakes operations.

\textbf{Use of Advanced Concepts:}

Identify the use of advanced programming concepts, design patterns, or architectural principles that indicate a higher difficulty level.

Assess whether the problem requires innovative solutions or custom implementations.

\textbf{Potential Risks and Challenges:}

Consider any risks associated with solving the problem, such as security concerns, data integrity issues, or compliance requirements.

Evaluate the likelihood of encountering unforeseen obstacles during development.

After analyzing these factors, provide an explanation for your estimation, highlighting the key indicators that influenced your determination of the problem's difficulty level.

Next, select the most appropriate difficulty level from the following options: \textbf{Very Easy, Easy, Moderate, Challenging, Very Challenging}

Then, generate a JSON object with the following structure:

\{

"commit\_hash": "insert commit hash here", 

"MP": "C",

"Q": "4",

"Answer": "insert estimated value that was selected earlier"

\}

\paragraph{Q5 Prompt}

Your task is to estimate the \textbf{maintainability} of the code changes in this commit. Based solely on the provided code, evaluate how easy it would be for developers to understand, modify, fix, and extend the code in the future.

To make a thorough and accurate assessment, please carefully consider the following aspects:

\textbf{Code Readability:}

Evaluate the clarity of the code, including naming conventions, code formatting, and use of comments where appropriate.

Assess whether the code is self-explanatory or if it requires additional documentation to be understood.

\textbf{Code Organization and Structure:}

Consider how the code is organized into classes, methods, and packages.

Evaluate the modularity and cohesiveness of the code components.

Determine if the code follows logical structuring that aids comprehension.

\textbf{Adherence to Coding Standards and Best Practices:}

Assess whether the code complies with industry-standard coding conventions and guidelines.

Evaluate the use of design principles such as the SOLID principles, DRY (Don't Repeat Yourself), and KISS (Keep It Simple, Stupid).

Consider the consistency of coding style throughout the changes.

\textbf{Complexity and Simplicity:}

Analyze the complexity of the code, including cyclomatic and n-path complexity and the use of complex algorithms or data structures.

Determine whether the code is as simple as possible while effectively achieving its purpose.

Identify any areas where unnecessary complexity may hinder understanding or maintenance.

\textbf{Use of Design Patterns and Reusability:}

Identify any design patterns implemented and evaluate whether they enhance maintainability.

Assess the reusability of code components and whether they are designed for easy extension or modification.

Consider how well the code anticipates future changes or additions.

\textbf{Error Handling and Robustness:}

Evaluate how the code handles exceptions, errors, and edge cases.

Assess the robustness of the code in dealing with unexpected inputs or conditions.

Determine whether proper error logging and reporting mechanisms are in place.

\textbf{Dependency Management:}

Assess how the code manages dependencies on external libraries, APIs, or modules.

Evaluate the impact of these dependencies on maintainability, including potential issues with updates or compatibility.

Consider whether the code abstracts dependencies appropriately to minimize coupling.

\textbf{Testability:}

Consider how easily the code can be tested, including the ability to write effective unit tests and integration tests.

Evaluate whether the code structure facilitates mocking or stubbing external dependencies.

Assess the presence of test hooks or interfaces that enhance testability.

\textbf{Documentation and Comments:}

\begin{itemize}
    \item Examine the presence and quality of inline comments and documentation.
    \item Determine whether complex sections of code are adequately explained.
    \item Consider the use of JavaDoc or other documentation practices to assist future developers.
\end{itemize}
After analyzing these factors, provide an explanation for your estimation, highlighting the key indicators that influenced your determination of the code's maintainability.

Next, select the most appropriate level of maintainability from the following options: \textbf{Poor, Below Average, Average, Good, Excellent}

Then, generate a JSON object with the following structure:

\{

"commit\_hash": "insert commit hash here", 

"MP": "C",

"Q": "5",

"Answer": "insert estimated value that was selected earlier"

\}

\end{document}